\documentclass[iop,apjl]{emulateapj}
\usepackage{color}
\newcommand{\widefig}[3]{%
\begin{figure*}[tbp]
\begin{center}
\includegraphics[width=180mm]{#1}
\caption{#3}
\label{#2}
\end{center}
\end{figure*}
}

\begin{document}

\title{First detection of sign-reversed linear polarization from the forbidden [O I] 630.03 nm line}
\shorttitle{Linear polarization from the [O I] 630.03~nm line}

\author{A.G.~de~Wijn}
\email{dwijn@ucar.edu}
\affil{High Altitude Observatory, National Center for Atmospheric Research, P.O. Box 3000, Boulder, CO 80307, USA}

\author{H.~Socas-Navarro \and\ N.~Vitas}
\affil{Instituto de Astrof\'\i{}sica de Canarias, Avda V\'\i{}a L\'actea S/N, La Laguna E-38205, Tenerife, Spain}

\shortauthors{De Wijn, Socas-Navarro, and Vitas}

\begin{abstract}
We report on the detection of linear polarization of the forbidden [O\,\textsc{i}] 630.03~nm spectral line.
The observations were carried out in the broader context of the determination of the solar oxygen abundance, an important problem in astrophysics that still remains unresolved.
We obtained spectro-polarimetric data of the forbidden [O\,\textsc{i}] line at 630.03~nm as well as other neighboring permitted lines with the Solar Optical Telescope of the Hinode satellite.
A novel averaging technique was used, yielding very high signal-to-noise ratios in excess of $10^5$.
We confirm that the linear polarization is sign-reversed compared to permitted lines as a result of the line being dominated by a magnetic dipole transition.
Our observations open a new window for solar oxygen abundance studies, offering an alternative method to disentangle the Ni\,\textsc{i} blend from the [O\,\textsc{i}] line at 630.03~nm that has the advantage of simple LTE formation physics.
\end{abstract}


\maketitle

\section{Introduction}\label{sec:introduction}

Forbidden lines have been extensively studied for solar coronal magnetometry, and successfully used to infer the orientation of the coronal magnetic field in the plane of the sky as well as the longitudinal strength \citep{2013SoPh..288..467J}.
However, forbidden lines are exceedingly weak when observed on the disk of the sun and are not commonly observed, with only a few exceptions.
One of these is the [O\,\textsc{i}] line at 630.03~nm.
We present here the first observations of linear polarization in this forbidden line and confirm that it exhibits the sign reversal compared to permitted lines, which is expected by theory.

Oxygen is the third most abundant element in the universe after hydrogen and helium.
It plays an important role in defining the structure of stellar interiors, both as an opacity source and as a donor of free electrons.
The abundance of other elements that also play an important role but do not have abundance indicators in the photospheric spectrum, such as Ne, is often measured in solar coronal lines relative to oxygen.
A precise knowledge of the oxygen abundance is crucial in, e.g., the construction of solar and stellar interior models that, in turn, are used for the development of stellar evolution theories and models and, ultimately, the dating of globular clusters and other astrophysical objects.
Unfortunately, our current understanding of the oxygen abundance in the solar system is neither complete nor accurate.
It is not possible to determine the oxygen abundance in meteorites because it is a highly volatile element.
The few spectral indicators oxygen produces in the solar photosphere are all either affected by complex formation physics, extremely weak forbidden transitions, and/or contaminated by blends.
The 630.03~nm line studied in this paper is both a forbidden transition and contaminated by a Ni\,\textsc{i} blend.

The old paradigm had an abundance of oxygen around 800~ppm determined first by \cite{1978MNRAS.182..249L} as an update to the result of \cite{1968MNRAS.138..143L}, and subsequently refined by others \citep[e.g.,][]{1989GeCoA..53..197A}.
The abundance of oxygen was determined by fitting one-dimensional semi-empirical models to spectral features produced by oxygen atoms and molecules that contain oxygen atoms.
The abundance of oxygen has been revised downward several times since \citep[e.g.,][]{1998SSRv...85..161G}.
Abundances under 500~ppm were proposed by \cite{2001ApJ...556L..63A} and \cite{2004A&A...417..751A} who accounted for the Ni blend, used updated atomic parameters, and a new generation of 3D models in their analyses.
The implications of adopting the lower value are so far-reaching across all fields of astrophysics that the ensuing controversy has been referred to in the literature as the \emph{solar oxygen crisis} \citep{2006ApJS..165..618A}.
More than a decade later, the solar oxygen abundance problem still awaits a satisfactory resolution.
Several papers from various groups have been published with disparate results \citep{2009ARA&A..47..481A}.
Recent work suggests that the systematic uncertainties in the traditional abundance determination analyses are much larger than previously thought, to the point that this much-debated discrepancy might be within the expected empirical error \citep{2015A&A...577A..25S}.
It is necessary to search for alternative diagnostics that could provide complementary information and break the current impasse.
Spectro-polarimetry is a novel observational technique in the context of abundance determinations that shows much promise \citep{2007ApJ...660L.153S,2008ApJ...682L..61C}.

\section{Observations and reduction}\label{sec:observations}

\widefig{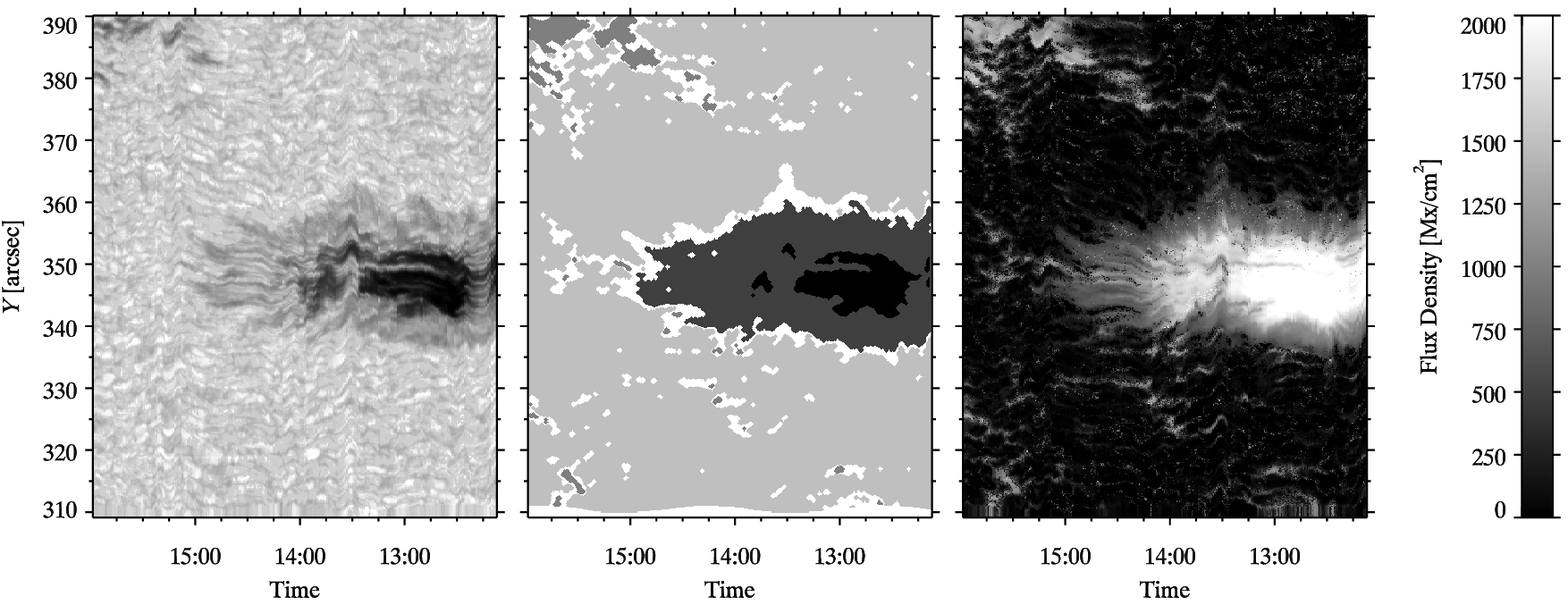}{fig:data_20101122}{\emph{Left to right:} Continuum intensity, area masks, and line-of-sight flux density scaled between $0$ and $2000~\mathrm{Mx}\,\mathrm{cm}^{-2}$.
The gray shading of the area masks corresponds from light to dark to weakly magnetized pixels, pixels in the magnetic network, plage, or pores, the sunspot penumbra, and sunspot umbra, respectively.
White pixels do not belong to any mask.}

The SpectroPolarimeter \citep[SP,][]{2013SoPh..283..579L} on the \emph{Solar Optical Telescope} \citep{2008SoPh..249..167T,2008SoPh..249..197S,2008SoPh..249..233I,2008SoPh..249..221S} of the Hinode spacecraft \citep{2007SoPh..243....3K} was used in an innovative mode to observe the spectral region from 629.97 to 630.35~nm.
The instrument normally observes a 0.24-nm wide region at 630.21~nm, but was programmed to observe the wider region split in two readouts in rapid succession with 10 pixels of overlap in the spectral direction.
The observations lasted for 3~hours and 52~minutes starting just after 12:07~UTC on November~22, 2010.
The slit was kept stationary and initially positioned at approx.\ $(5\arcsec,350\arcsec)$, slightly ahead of a sunspot that then moved through the slit aperture due to solar rotation.
An area of approx.\ $36\arcsec\times82\arcsec$ was rastered in this way.
It is shown in Fig.~\ref{fig:data_20101122}.
The image shows substantial distortion that results from pointing drift that is corrected by the correlation tracker under normal operating conditions.
In this case, however, the correlation tracker must be reset after each pair of exposures in order to raster the FOV because it tracks solar rotation as well as pointing drift.

The two readouts were first merged and then processed with the standard calibration code that was modified to accept the wider spectral region.
The calibrated data was then processed with the NCAR/HAO MERLIN Milne-Eddington inversion code\footnote{\url{http://www2.hao.ucar.edu/csac/csac-spectral-line-inversions}} to yield the flux density and the orientation of photospheric magnetic field from the two Fe\,\textsc{i} lines at 630.15 and 630.25~nm.

Masks were created to separate weakly-magnetized areas from the sunspot umbra and penumbra, and from areas with strong magnetic field outside of the sunspot.
The weakly-magnetized area mask is defined as those pixels that have a flux density below $400~\mathrm{Mx}/\mathrm{cm}^2$, while the strong-field area mask contains those pixels that have flux density above $600~\mathrm{Mx}/\mathrm{cm}^2$.
The umbra of the sunspot is defined as the pixels in the strong-field area mask that have a continuum intensity less than half the mean continuum intensity of the field of view.
The masks are filtered to first remove structures smaller than a diamond shape of $5\times5$ pixels, then filtered to fill in holes smaller than that diamond shape.
The penumbral area mask is then created with those pixels in the strong-field area mask that are part of the features that overlap with the umbral area mask, but that are not in the umbral area mask.
Finally, the data was filtered to remove noise using a low-pass filter in the spectral domain.

The [O\,\textsc{i}] line that is the subject of our study is so weak that its linear polarization signal is only similar in magnitude to the photon noise in the strongly magnetized environment of a sunspot.
In principle, it is possible to improve the signal-to-noise (SNR) ratio of the observations at the cost of sacrificing spatial resolution by averaging over the field of view.
However, this is not straightforward in the case of polarization profiles whose shape and sign depend on the geometry of the magnetic field.
Linear polarization signals emerging from magnetic regions that have a 90-degree azimuth separation in the frame of the observer are opposite and cancel each other.

The magnetic field inferred from the Fe\,\textsc{i} lines is likely to be slightly biased toward higher layers than those probed by the [O\,\textsc{i}] line, but they are both formed in the photosphere and the variation with height of the field orientation is expected to be small.
We thus use the magnetic field geometry derived from the inversion of the magnetically sensitive Fe\,\textsc{i} lines, observed simultaneously and cospatially, to rotate the Stokes Q and U profiles at each pixel to a common reference frame before averaging.
Differences in formation height would merely introduce a small amount of signal cancellation in our procedure, making the profile detection harder but not invalidating the results presented below.

\widefig{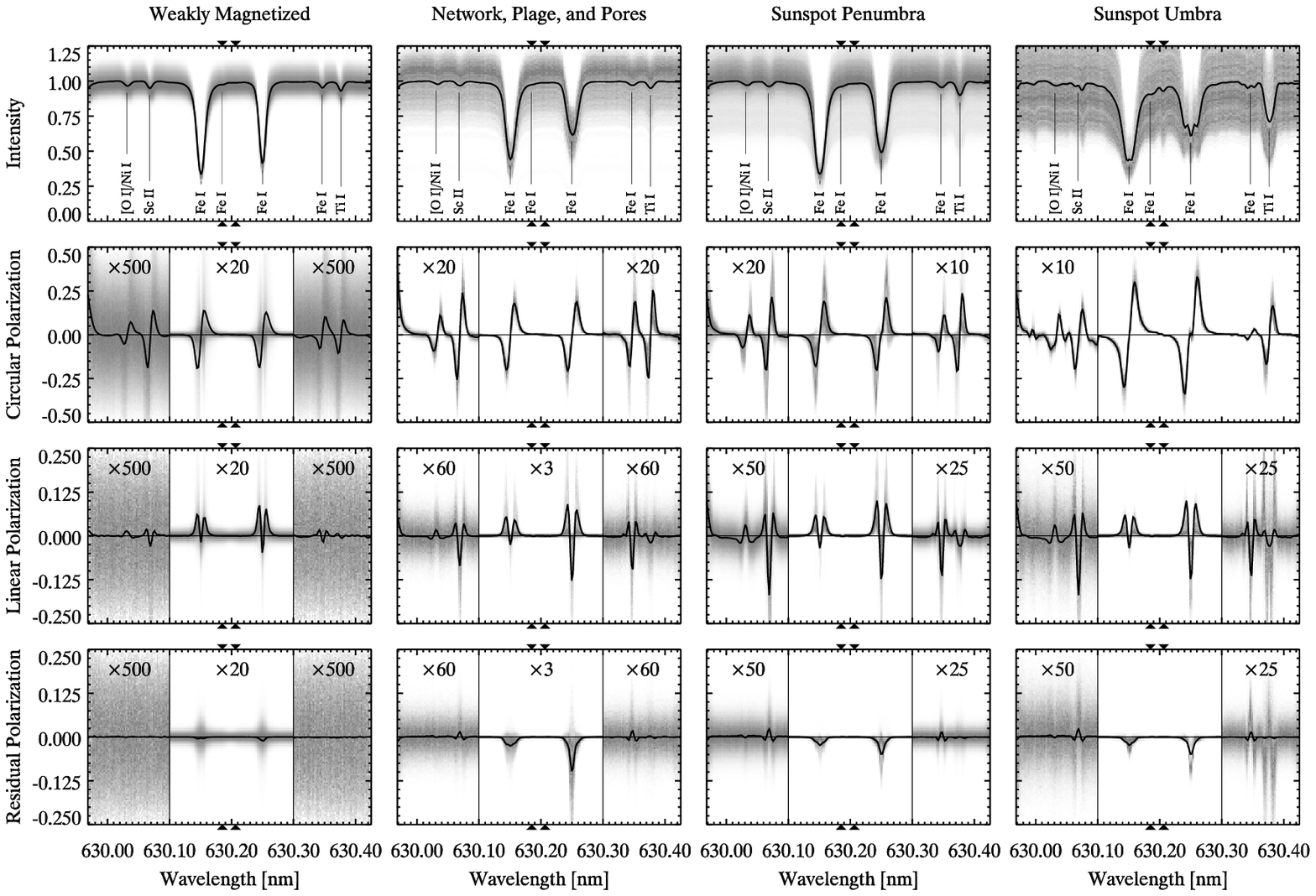}{fig:avg_profiles}{Average intensity and polarization profiles in the areas defined in Fig.~\ref{fig:data_20101122}.
The gray shade background shows the spread of profiles.
Prominent lines are identified in the intensity plots.
Some parts of the profiles have been enlarged to show details.
Residual polarization refers to what is left in the Stokes-U profile after the profiles have been rotated to a reference frame in which all of the signal should be in the Stokes-Q profile.
The arrowheads at the top and bottom of each panel indicate the area of overlap between the two detector readouts.}

We can estimate the resulting SNR by the propagation of random error.
The masks for the weakly magnetized areas, strong fields, penumbra, and umbra contain 150140, 36833, 26376, and 5612 pixels, respectively.
The SNR of a standard observation using both beams of the polarization analysis and a $4.8~\mathrm{s}$ exposure was approx.~$10^3$ toward the end of 2010 \citep{2013SoPh..283..579L}.
Our observations use both beams but have a longer exposure time of $12.8~\mathrm{s}.$
We thus find SNRs of $6.3\times10^5$, $3.1\times10^5$, $2.6\times10^5$, and $1.2\times10^5$, for the spectra averaged over all pixels in the above masks.
Polarimetry relies on a measurement of differences, and is consequently substantially less sensitive to systematic errors than a measurement of intensity.
With such high SNR it is however reasonable to assume that systematic errors will dominate even if they are not readily apparent.

\section{Discussion}\label{sec:discussion}

The observed spectral range shown in Fig.~\ref{fig:avg_profiles} contains six prominent spectral lines, with a seventh just at the blue extreme of the spectral FOV.
All six lines exhibit the same sign pattern in circular polarization (negative-positive from blue to red).
In linear polarization, however, we observe a reversal of the polarization signal.
All but the [O\,\textsc{i}] line show the same pattern (positive-negative-positive), while that line exhibits just the opposite behavior (negative-positive-negative).
In other words, the linear polarization signal in this line has the opposite sign compared to the five others in the observed spectral range.
All of these spectral features have been observed simultaneously and have gone through the same calibration and data reduction procedures.
The effect is present in all areas defined in Fig.~\ref{fig:data_20101122}.
The signal is predictably weakest in the weakly-magnetized areas, but still visisble.
While some of the lines are much stronger (e.g., the Fe\,\textsc{i} 630.15 and 630.25~nm lines), others are of similar strength (e.g., the Sc\,\textsc{ii} 630.07~nm line), which rules out reasons for the difference such as a specific magnetic field configuration.

The pertinent difference between the [O\,\textsc{i}] line and all other transitions in the spectra is that its electric dipole component is forbidden by quantum selection rules.
The absorption is dominated by the much weaker magnetic dipole component.
The observed polarization profile is in agreement with the theory of generation of polarized light that predicts that the linear polarization originated by a magnetic dipole term has the opposite sign structure to that produced by the electric dipole term \citep[][Sect.~6.8]{2004ASSL..307.....L}.
The circular polarization, on the other hand, has the same sign in the magnetic and the electric dipole terms.

We observe that the linear polarization signal of the forbidden [O\,\textsc{i}] line dominates that of the permitted Ni\,\textsc{i} line.
This is in agreement with expectations as the [O\,\textsc{i}] line is stronger than the Ni\,\textsc{i} line even when assuming a very low oxygen abundance \citep{2001ApJ...556L..63A}, at least outside the sunspot.
Furthermore, the effective Land\'e factor for the [O\,\textsc{i}] line is higher than for the Ni\,\textsc{i} line \citep[1.25 and 0.51, respectively,][]{2008ApJ...682L..61C}.

A precise understanding of the [O\,\textsc{i}] line polarization will open new windows for the challenging diagnostics of the solar oxygen abundance.
The blended Ni\,\textsc{i} line is an electric dipole transition.
Therefore, the [O\,\textsc{i}] and Ni\,\textsc{i} lines exhibit similar behavior in Stokes I and V, but opposite in Stokes Q and U.
The intensity spectrum of the blend will remain similar to a single line because the blend is unresolved, i.e., the line widths are greater than the separation of the components.
However, the small wavelength shift between the lines will give rise to an intricate pattern of features in the polarized spectra that can help resolve ambiguities because the polarized line profiles have structure on smaller scales than the intensity spectrum \citep[see][]{2008ApJ...682L..61C}.
The different solar features in the FOV have different thermal and magnetic properties that will affect the [O\,\textsc{i}] and Ni\,\textsc{i} lines differently, and thus result in different, distinct patterns.
Measurements of the polarization properties of the line can yield a powerful handle on relative strengths that should allow us to determine accurate constraints on the relative abundances of O and Ni.
However, the determination of the oxygen abundance will require careful consideration of the many intricacies of the formation of this closely blended line, as well as attention to oxide formation particularly in the umbra, even if the relative strengths of the components are well determined.

\section{Conclusion}\label{sec:conclusion}

We have presented the first direct comparison of simultaneously observed linear polarization in forbidden and permitted lines formed in the solar photosphere.
Our observation of the linear polarization profile of the [O\,\textsc{i}] line has been made possible due to a novel averaging method of Stokes profiles that takes the signal polarity into account.
The observed behavior is in agreement with the theory of generation of polarized light, and, in the context of the solar oxygen abundance, carries the promise of a robust way to differentiate the [O\,\textsc{i}] line at $630~\textrm{nm}$ from the Ni\,\textsc{i} blend that historically has compromised the desirable utility of the LTE formation of the forbidden oxygen line.

\acknowledgments
The National Center for Atmospheric Research is sponsored by the National Science Foundation.
HSN and NV gratefully acknowledge support from the Spanish Ministry of Economy and Competitivity through project AYA2014-60476-P (Solar Magnetometry in the Era of Large Solar Telescopes).
AdW is grateful to the Spanish MINECO for supporting a visit to the IAC through the 2011 Severo Ochoa Program SEV-2011-0187.
The data reported in this paper are available from the Hinode Data Archive (\url{http://hinode.nao.ac.jp/hsc\_e/darts\_e.shtml}).

\end{document}